\documentclass[12pt]{article}%
\usepackage{amsmath}
\usepackage{amsfonts}
\usepackage{amssymb}
\usepackage{graphicx}%
\providecommand{\U}[1]{\protect\rule{.1in}{.1in}}

\begin{document}

\title{The Friedel-Anderson and Kondo Impurity  Problem for Arbitrary s-Band Density
of States and Exchange Interaction.}
\author{Gerd Bergmann\\Department of Physics\\University of Southern California\\Los Angeles, California 90089-0484\\e-mail: bergmann@usc.edu}
\date{\today}
\maketitle

\begin{abstract}
In his renormalization paper of the Kondo effect Wilson replaced the full band
of s-electrons by a small number of "Wilson states". He started from a rather
artificial symmetric band with constant density of states and constant
interaction with the impurity. It is shown in the present paper that with a
minor modification the Wilson states are optimally suited to treat the
interaction of an impurity with an arbitrary s-band. Each Wilson state
represents electrons of a whole energy range $\Delta\varepsilon$. It carries
the interaction of all these electrons with the impurity. All the other
electron states in this energy range have zero interaction with the impurity
and are neglected in the calculation. The resulting error is minor. As an
example the singlet-triplet excitation energy of a Kondo impurity is
numerically calculated for a tight-binding band with a strongly energy
dependent density of states.


\newpage

\end{abstract}

\section{Introduction}

The description of metal electrons as plane waves $\psi_{\mathbf{k}}\left(
\mathbf{r}\right)  =e^{i\mathbf{kr}}/\sqrt{V}$ has been extremely successful
during the past. However, if one wants to manipulate the whole electron basis
one runs into problems. For example, a rotation of the whole basis in Hilbert
space requires a number of Euler angles which is equal to $N\left(
N-1\right)  /2$ where $N$ is the number of basis states. Such a rotation is
rather difficult for $N=10^{23}$ in a macroscopic metal sample. Wilson
\cite{W18} encountered a similar problem when he treated the Kondo effect
where the impurity mixes the plane waves in complex ways. For his numerical
evaluation Wilson strongly reduced the number of states to a few states and
only included these "Wilson"-states in his treatment. (Wilson called these
states "Kondo"-states).

The purpose of this paper is to illuminate some of the physics behind the use
of the Wilson states and to introduce a generalization. We will consider a
host with a d-impurity. Wilson used a band of electrons with constant density
of states and a constant interaction with the impurity. This may appear to be
rather artificial and far removed from a real system. This is far from true.
With a minor modification the Wilson states can be applied to bands with
arbitrary energy density of states and impurity interaction. This
generalization does not increase the complexity of the calculation.

To demonstrate the method two examples will be treated quantitatively in this
paper. Our group developed recently a new and very compact approximate
solution for the Friedel-Anderson and the Kondo impurity problems \cite{B151},
\cite{B152}, \cite{B153}. This method yields remarkably accurate results for
the ground-state energy and the singlet-triplet excitation energy. The ground
state requires the construction of one localized s-electron state (per spin)
$a_{0}^{\dag}=%
{\textstyle\sum}
\alpha_{0}^{\mathbf{k}}c_{\mathbf{k}}^{\dag}$ while the remaining electron
states are rearranged into states $a_{i}^{\dag}$ so that the new basis
$\left\{  a_{i}^{\dag}\right\}  $ is orthonormal. Since this requires an
orthogonal transformation of the whole electron basis Wilson's state reduction
is essential in the numerical procedure. In section IV the ground-state energy
and the singlet-triplet excitation energy for a Kondo impurity is numerically
calculated for (i) a free electron system with $\rho\left(  \varepsilon
\right)  \varpropto\sqrt{\varepsilon}$ and (ii) a tight-binding metal. For
both electron systems the density of states is not constant. Particularly in
the tight binding band $\rho\left(  \varepsilon\right)  $ show a strong
variation (see Fig.2).

\newpage

\section{The Wilson States}

Let us first start (like Wilson) with the somewhat artificial energy band
having the half-band width $D$, i.e., the range $\left[  -D:D\right]  $ with a
constant density of states. It is half filled and the Fermi level lies at zero
energy. We divide all energies by the half-width $D$ of the band (see Fig.1).
Then the band extends from $-1$ to $+1$. The full band can accommodate one
electron per atomic volume (and spin). The density of states (per spin) is
$\rho_{0}=1/2$ per atomic volume $v_{A}$.

Wilson sub-divided the positive and negative parts of the energy band into $N$
cells. This is done on an exponential scale. The boundaries of the cells are
for $\nu<N/2$ at $\xi_{\nu}=-1/2^{\nu}$, i.e. $-1,-1/2,..-1/2^{2},....$ and
for $\nu>N/2$ at $\xi_{\nu}=+1/2^{N-\nu}$, i.e. ...,$1/2^{2},1/2$, $1$.
(Wilson used originally $\pm1/\Lambda^{\nu}$ with the parameter $\Lambda$ but
generally chose the value $\Lambda=2$). With these discrete points $\xi_{\nu}$
one obtains a sequence of intervals or cells. The interval $\nu$ (for $\nu$%
$<$%
$N/2$) is given by $\xi_{\nu-1}=-1/2^{\nu-1}<$ $\varepsilon$ $<-1/2^{\nu}%
=\xi_{\nu}$ or $\left[  \xi_{\nu-1}:\xi_{\nu}\right]  $. (The equivalent
intervals for positive $\xi$-values are treated completely analogously).

The new Wilson states $\psi_{\nu}$ lie in the center of the energy interval
$\left(  \xi_{\nu-1},\xi_{\nu}\right)  $ and have an energy of $\varepsilon
_{\nu}=\left(  \xi_{\nu}+\xi_{\nu-1}\right)  /2$, i.e. $-\frac{3}{4},-\frac
{3}{8},-\frac{3}{16},..,-\frac{3}{2\cdot2^{N/2}},-\frac{1}{2\cdot2^{N/2}}$
(for $\nu\leq N/2$). One essential property of the Wilson basis is that it has
an arbitrarily fine energy spacing at the Fermi energy. In his normalization
treatment of the Kondo effect Wilson entangled these states numerical with the
Kondo impurity.

The first cell $\mathfrak{C}_{1}$ extends from $\left(  -1:-1/2\right)  .$
This cell contains $Z_{1}$($\thickapprox10^{22}$) electron states which we
denote as $\varphi_{\mu}\left(  \mathbf{r}\right)  $. They all interact with
the impurity. Now these states are transformed with the matrix $\mathbf{A}$
where $\mathbf{A}$ has the simple form $A_{\lambda\mu}=\exp\left(
i2\pi\lambda\mu/Z_{1}\right)  /\sqrt{Z_{1}}$. With $A_{\lambda\mu}$ one
obtains the new states%

\[
\chi_{\lambda}\left(  \mathbf{r}\right)  =\frac{1}{\sqrt{Z_{\nu}}}%
{\textstyle\sum}
\varphi_{\mu}\left(  \mathbf{r}\right)  e^{i2\pi\lambda\mu/Z_{\nu}}%
\]
where $0\leq\lambda<Z_{1}$ and the sum is taken over all states $\varphi_{\mu
}\left(  \mathbf{r}\right)  $ in $\mathfrak{C}_{1}$.%

\begin{align*}
&
{\includegraphics[
height=3.4388in,
width=4.2507in
]%
{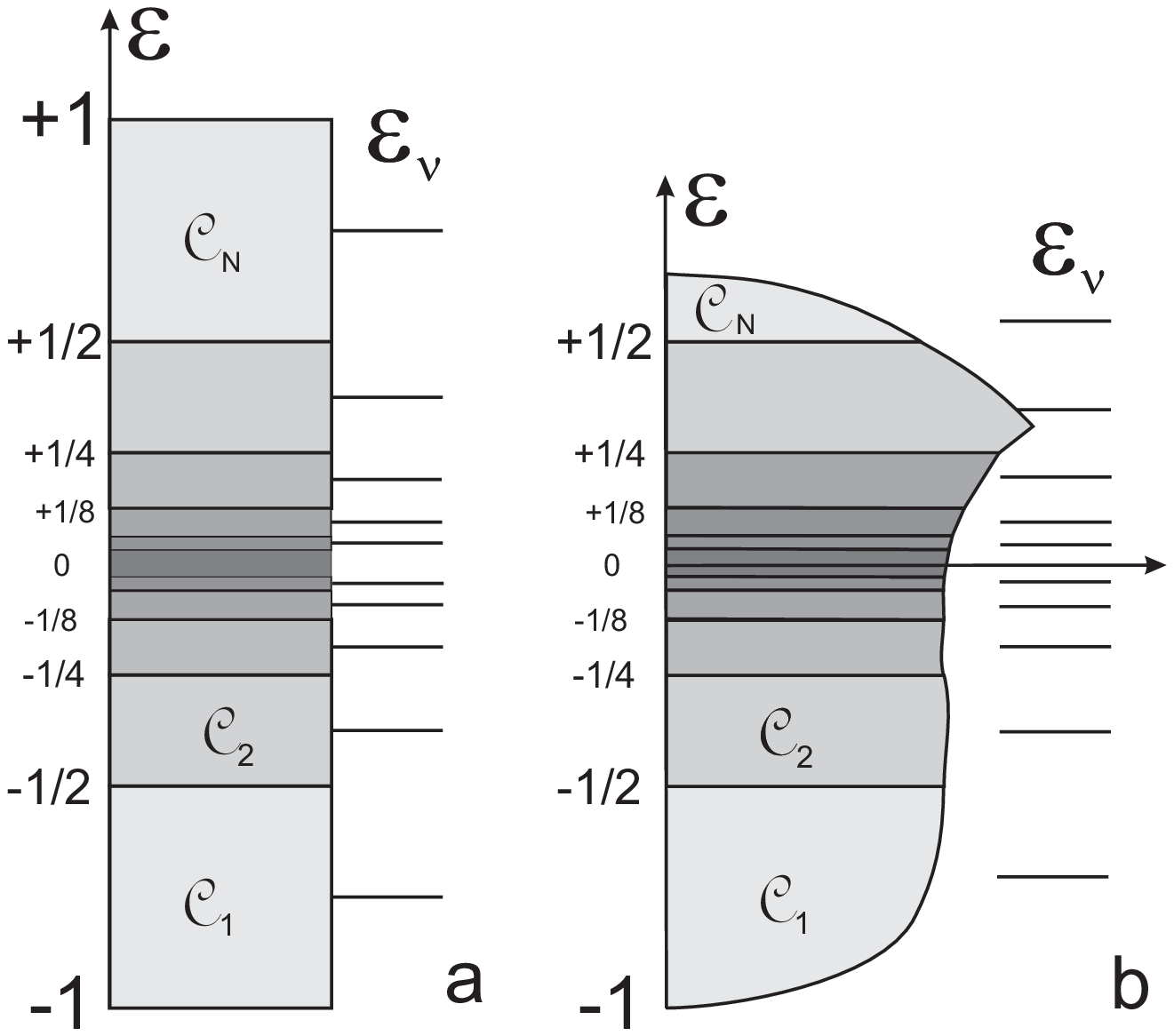}%
}%
\\
&
\begin{tabular}
[c]{l}%
Fig.1a: Left side: Wilson's rectangular symmetric energy band.\\
It is divided into energy cells whose width decrease\\
exponentially towards the Fermi energy. Each energy cell $\nu$\\
is represented by a state $\psi_{\nu}$ whose energy lies in the center\\
of the cell. It represents the full interaction of all states in the cell\\
with the impurity.\\
Fig.1b: A realistic band with energy dependent density of states.\\
The division into energy cell is completely analogous. The\\
construction of the Wilson states is modified to accommodate\\
the energy dependence of density of state and s-d-interaction.
\end{tabular}
\end{align*}

If all states $\varphi_{\mu}\left(  \mathbf{r}\right)  $ have the same s-d
matrix element $v_{sd}$ with the impurity then only the state for $\lambda=0$
has a finite interaction with the impurity. For all other states with
$\lambda\neq0$ the s-d interaction is zero. These states are ignored in the
calculation. Then the cell $\mathfrak{C}_{1}$ is represented by a single state
$\psi_{1}\left(  \mathbf{r}\right)  $ where%

\begin{equation}
\psi_{1}\left(  \mathbf{r}\right)  =%
{\textstyle\sum}
\varphi_{\mu}\left(  \mathbf{r}\right)  /\sqrt{Z_{1}} \label{psi_1}%
\end{equation}
The energy (expectation value) of this state follows from the energies
$e_{\mu}$ of the states $\varphi_{\mu}$
\[
\varepsilon_{1}=%
{\textstyle\sum}
e_{\mu}/Z_{1}=\frac{1}{2}\left(  \xi_{0}+\xi_{1}\right)
\]
This state has the s-d-matrix element $V_{sd}\left(  1\right)  =\sqrt{Z_{1}%
}v_{sd}$.

The same procedure is then applied to all the other cells $\mathfrak{C}_{\nu}%
$. In other words each energy cell is represented by just one state $\psi
_{\nu}$ which carries the full interaction with the impurity. The neglected
states have no interaction with the impurity.

For the discussed case (constant density of state and constant interaction)
the s-d-matrix elements $V_{sd}\left(  \nu\right)  $ of the Wilson states are
proportional to $\sqrt{Z_{\nu}}$ and given by
\[
V_{sd}\left(  \nu\right)  =V_{sd}^{0}\sqrt{\frac{\Delta\xi_{\nu}}{2}}%
\]
where $\Delta\xi_{\nu}=\xi_{\nu}-\xi_{\nu-1}$ is the width of the energy cell
and $\left\vert V_{sd}^{0}\right\vert ^{2}=%
{\textstyle\sum_{\nu}}
\left\vert V_{sd}\left(  \nu\right)  \right\vert ^{2}$ $=%
{\textstyle\sum_{\mathbf{k}}}
\left\vert v_{sd}\right\vert ^{2}$

The size of the energy cells is about equal to the distance from the Fermi
level. From perturbation theory one realizes that each Wilson state has about
the same coupling to the impurity. (The ratio of the square of the matrix
element divided by the energy denominator $\left\vert V_{sd}\left(
\nu\right)  \right\vert ^{2}/\Delta\varepsilon_{\nu}$ yields about the same
contribution as long as $\Delta\varepsilon_{\nu}$ can be replaced by
$\varepsilon_{\nu}$ itself.)

Wilson's philosophy was somewhat different when he introduced the
"Kondo"-states. In his renormalization approach he considered the wave
functions of these states as spherical shells, like in an onion, surrounding
the impurity. The states $\psi_{\nu}$,$\psi_{N+1-\nu}$ belonged to the
$\left(  \nu-1\right)  $-th shell. During the renormalization procedure he
first coupled the states $\psi_{1},\psi_{N}$ directly to the impurity. Then he
progressed step wise outwards by coupling the impurity complex (consisting of
the impurity plus $\left(  \nu-1\right)  $ shells) to the shell $\nu$. In our
pragmatic application we consider the Wilson states as representatives of a
finite energy range which contain the complete s-d-interaction of all the
(neglected) states in the energy range with the impurity. The square of the
resulting $V_{sd}\left(  \nu\right)  $ is proportional to the total number of
states in that energy interval.

The neglected states which do not interact with the impurity cause a small
error since they interact through the kinetic energy with the states
$\psi_{\nu}$. But Wilson pointed out that this error is very small, a result
that our calculations confirm.

\newpage

\section{Generalization of Wilson States}

The Wilson concept can now be applied to an s-band with arbitrary density of
states and state-dependent s-d-interaction. Numerically it is a trivial
extension of the Wilson rectangular band. Let us consider a band that extends
from $-D_{1}$ to $+D_{2}$ with the Fermi energy at zero energy. The density of
states depends strongly on the energy and the matrix elements for the
s-d-interaction $v_{sd}$ are not only a function of the energy but depend on
the exact quantum numbers of the state. To simply the discussion we normalize
the energy with $D=\max\left(  D_{1},D_{2}\right)  $ so that the band lies
within $\left[  -1:+1\right]  $ (see fig.1b). Then we subdivide the band as
before obtaining the same energy cells $\mathfrak{C}_{\nu}$. Again we use the
cell $\mathfrak{C}_{1}$ to demonstrate the new procedure. This cell contains
$Z_{1}$ electron states $\varphi_{\mu}\left(  \mathbf{r}\right)  $.

\subsection{Friedel-Anderson impurity}

Let us first consider the case where the electrons interact with the impurity
through the s-d-interaction. $\ $Now this interaction $v_{sd,\mu}$ with the
impurity depends on the state $\varphi_{\mu}$. In this case we define the
Wilson state $\psi_{1}\left(  \mathbf{r}\right)  $ as%

\begin{equation}
\psi_{1}\left(  \mathbf{r}\right)  =%
{\textstyle\sum}
v_{sd,\mu}\varphi_{\mu}\left(  \mathbf{r}\right)  /\sqrt{%
{\textstyle\sum}
\left\vert v_{sd,\mu}\right\vert ^{2}} \label{psi_1'}%
\end{equation}
(The sum over $\mu$ is taken over all states $\varphi_{\mu}$ within the energy
cell $\mathfrak{C}_{1}$.) The energy $\varepsilon_{1}$ of this state $\psi
_{1}$ follows from the energies $e_{\mu}$ of the states $\varphi_{\mu}$
\begin{equation}
\varepsilon_{1}=\frac{%
{\textstyle\sum}
\left\vert v_{sd,\mu}\right\vert ^{2}e_{\mu}}{%
{\textstyle\sum}
\left\vert v_{sd,\mu}\right\vert ^{2}} \label{eps_1}%
\end{equation}

This state has the full s-d-interaction of the cell $\mathfrak{C}_{1}$ with
impurity. All other $\left(  Z_{1}-1\right)  $ states which can be constructed
from the states $\varphi_{\mu}$within cell $\mathfrak{C}_{1}$ (being
orthonormal to each other and to $\psi_{1}\left(  \mathbf{r}\right)  )$ have
zero interaction with the impurity. This can be easily seen with the following
argument. The state $\psi_{1}$ has the coefficients $\alpha_{1}^{\mu
}=v_{sd,\mu}/\sqrt{%
{\textstyle\sum}
\left\vert v_{sd,\mu}\right\vert ^{2}}$ in the basis of $\left\{  \varphi
_{\mu}\right\}  $. If $\chi=%
{\textstyle\sum}
\beta^{\mu}\varphi_{\mu}$ is a state orthogonal to $\psi_{1}$ (built from the
same basis $\left\{  \varphi_{\mu}\right\}  $) then the scalar product
$\left\langle \chi|\psi_{1}\right\rangle $ has to be zero, i.e.
\[
\left\langle \chi|\psi_{1}\right\rangle =%
{\textstyle\sum}
\alpha_{1}^{\mu}\beta^{\mu}=\frac{1}{\sqrt{%
{\textstyle\sum}
\left\vert v_{sd,\mu}\right\vert ^{2}}}%
{\textstyle\sum}
v_{sd,\mu}\beta^{\mu}=0
\]
The s-d-matrix element between the state $\chi$ and the impurity is equal to
the sum of its components, i.e.
\[
V_{sd}\left(  \chi\right)  =%
{\textstyle\sum}
v_{sd,\mu}\beta^{\mu}=0
\]
From the orthogonality with respect to the state $\psi_{1}$ follows the
disappearance of the interaction. Therefore all other states built from the
$\left\{  \varphi_{\mu}\right\}  $-basis have no interaction with the impurity.

So now we have a new Wilson basis $\left\{  \psi_{\nu}\right\}  $ with the
energies $\varepsilon_{\nu}$ and the s-d-interaction $V_{sd}\left(
\nu\right)  $. The energies $\varepsilon_{\nu}$ don't lie anymore in the
center of the cells and the s-d-matrix elements are no longer proportional to
the square root of the cell width. (There may be even a cell or two with no
Wilson state in it. Then one can either reduce the number of Wilson states or
one gives these states zero s-d-interaction.) For a numerical evaluation this
generalization of the Wilson states to arbitrary density of states and
interaction causes no additional complications.

\subsection{Kondo impurity}

For a Kondo impurity the situation is slightly more difficult because the
interaction matrix element $\left\langle \varphi_{\mu}\left(  \mathbf{r}%
\right)  \left\vert J\left(  \mathbf{r}\right)  \right\vert \varphi
_{\mu^{\prime}}\left(  \mathbf{r}\right)  \right\rangle $ is taken between two
s-electrons. We restrict ourselves here to the case where the exchange
potential is a $\delta$-function $J\left(  \mathbf{r}\right)  =J\delta\left(
\mathbf{r}\right)  $. In this case the matrix element takes the value
$J\varphi_{\mu}^{\ast}\left(  0\right)  \varphi_{\mu^{\prime}}\left(
0\right)  $ and can be separated. Therefore the role of\ $v_{sd,\mu}$ in
equ.(\ref{psi_1'}) is replaced by $\varphi_{\mu}\left(  0\right)  $. Then the
Wilson state $\psi_{1}$ for the Kondo impurity is
\begin{equation}
\psi_{1}\left(  \mathbf{r}\right)  =%
{\textstyle\sum}
\varphi_{\mu}\left(  0\right)  \varphi_{\mu}\left(  \mathbf{r}\right)  /\sqrt{%
{\textstyle\sum}
\left\vert \varphi_{\mu}\left(  0\right)  \right\vert ^{2}} \label{psi_1''}%
\end{equation}

These Wilson states have a maximal amplitude at the origin. All other states
in $\mathfrak{C}_{1}$ vanish at the origin and therefore they don't interact
with the $\delta$-exchange potential.

The matrix element between two Wilson states is
\begin{equation}
J_{\nu,\nu^{\prime}}=J\sqrt{%
{\textstyle\sum_{\mu\ni\mathfrak{C}_{\nu}}}
\left\vert \varphi_{\mu}\left(  0\right)  \right\vert ^{2}}\sqrt{%
{\textstyle\sum_{\mu\ni\mathfrak{C}_{\nu^{\prime}}}}
\left\vert \varphi_{\mu}\left(  0\right)  \right\vert ^{2}} \label{J_nn'}%
\end{equation}

This means that treatment of the Friedel-Anderson and the Kondo impurity can
be directly combined, for example, with density functional theory calculations
of the s-electron band, the Coulomb and s-d-exchange interaction. One has just
to extract from the band-structure calculation the energies and
matrix-elements of a relatively small number $N$ of states (where $N$ is of
the order 40)

\newpage

\section{Application of the Modified Wilson States in the Kondo Impurity
Problem}

As an example I calculate the Kondo ground state and its singlet-triplet
excitation energy for three examples and compare the results. For the
calculation I choose (i) the Wilson spectrum with $\rho\left(  \varepsilon
\right)  $=$1/2$=const, (ii) a three-dimensional free electron band with
$\rho\left(  \varepsilon\right)  \varpropto\sqrt{\varepsilon}$ and (iii) a
tight-binding band for an fcc lattice. The density of states of the latter is
sketched in Fig.3. These calculations are an extension of a recent paper on
the Kondo effect \cite{B153}. In the appendix the authors ansatz for the Kondo
solution is briefly sketched. It is discussed in detail in \cite{B153}.%

\begin{align*}
&
{\includegraphics[
height=2.6177in,
width=2.8335in
]%
{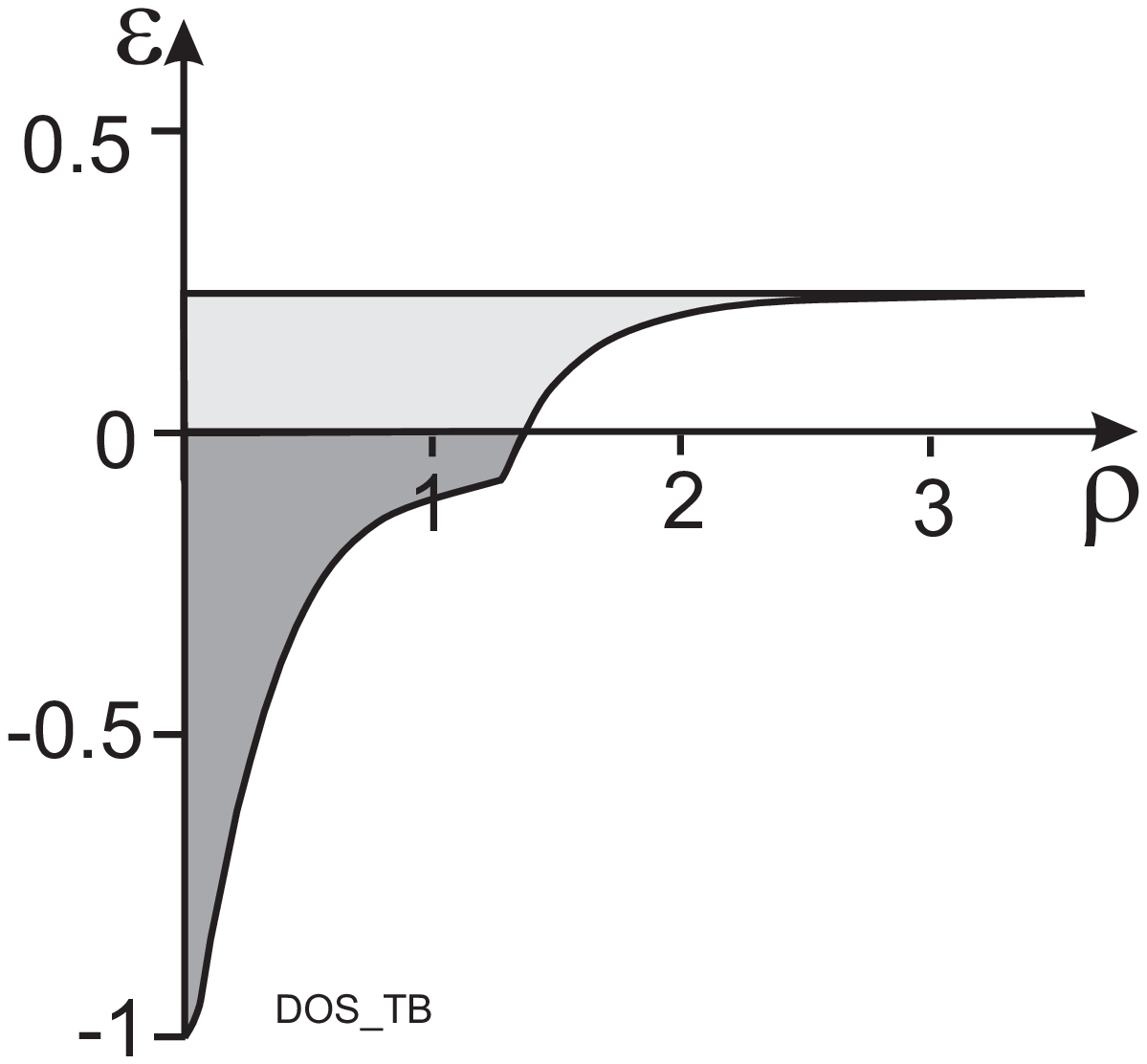}%
}%
\\
&
\begin{tabular}
[c]{l}%
Fig.2: Energy band for an fcc lattice in the\\
tight binding model. The dark shaded region\\
represents the half-filled band.
\end{tabular}
\end{align*}

\subsection{Comparison between the different band-structures.}

\subsubsection{Rectangular Wilson band}

Wilson's original rectangular band, extending from $\left(  -1:+1\right)  $
with $\rho_{0}=1/2,$ is used. The singlet-triplet excitation energy is plotted
in Fig.3a as a function of $1/\left(  2\rho_{0}J\right)  $. It follows
essentially an exponential law: $\Delta E_{st}\varpropto\exp\left[  -1/\left(
2\rho_{0}J\right)  \right]  .$

\subsubsection{Free electron band.}

Here the s-electrons are treated as free electrons. However, the s-band is
defined as full (i.e. cut off) at the energy where the electron density of one
electron per spin and atom is reached. Then the s-band is occupied with 1/2
electron for spin up and down. This defines the Fermi energy which can be
expressed in terms of the atomic volume
\[
\varepsilon_{F}=\frac{\hbar^{2}}{2m}\left(  \frac{3\pi^{2}}{v_{a}}\right)
^{2/3}%
\]
When the energy is normalized by the Fermi energy and the Fermi level adjusted
to zero the upper band edge lies at $2^{2/3}-1\thickapprox\allowbreak
0.587\,4.$ The density of states at the Fermi energy per atom and spin is
$\rho_{0}=3/4.$ The singlet-triplet excitation energy is plotted in Fig.2 as
stars. The (almost) straight line lies below the results of the rectangular
Wilson band but is parallel to it.

\subsubsection{Three dimensional tight binding}

For a face centered cubic metal the tight binding energy is given by%
\[
\varepsilon\left(  \mathbf{k}\right)  =E_{0}-4t\left[
\begin{array}
[c]{c}%
\cos\left(  \frac{1}{2}k_{x}a\right)  \cos\left(  \frac{1}{2}k_{y}a\right)  \\
+\cos\left(  \frac{1}{2}k_{y}a\right)  \cos\left(  \frac{1}{2}k_{z}a\right)
+\cos\left(  \frac{1}{2}k_{z}a\right)  \cos\left(  \frac{1}{2}k_{x}a\right)
\end{array}
\right]
\]
where $E_{0}$ is a constant energy, $t$ is the coupling between nearest
neighbors, i.e., the hopping matrix element, $a$ is the cubic lattice constant
and $\mathbf{k=}\left(  k_{x},k_{y},k_{z}\right)  $ is the wave number of the
electron. By calculating the energy for all k-states in the first Brillouin
zone (for a crystal of roughly $10^{6}$ atoms or $\mathbf{k}$-points in the
1BZ) one obtains a density of state which, after some smoothing, is shown in
Fig.2. The actual values of $E_{0}$ and $t$ cancel out through the
normalization with the Fermi energy. The density of states per spin and atom
at the Fermi energy is $\rho_{0}=1.36$ and is much larger than in the
rectangular Wilson band. The singlet-triplet excitation energy is plotted in
Fig.3 as triangles. The (almost) straight line lies below the results of the
rectangular Wilson and the free electron band but is parallel to both.

All three curves follow essentially the exponential law $\Delta E_{st}%
/D\varpropto$ $\exp\left[  -1/\left(  2J\rho_{0}\right)  \right]  $. For the
half-band width $D$ the value $D=1$ is used in all three curves.%

\begin{align*}
&
{\includegraphics[
height=3.3566in,
width=3.9219in
]%
{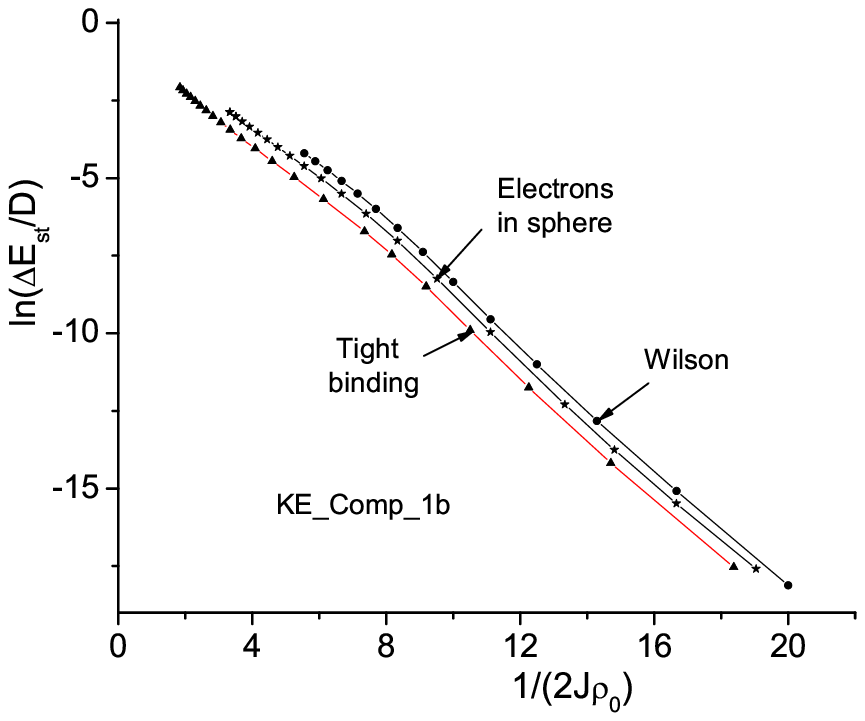}%
}%
\\
&
\begin{tabular}
[c]{l}%
Fig.3a: The logarithm of the singlet-triplet excitation energy\\
$\ln\left(  \Delta E_{st}/D\right)  $ is plotted as a function of $1/\left(
2J\rho_{0}\right)  $ for three\\
different spectra with different density of states $\rho_{0}$ at the\\
Fermi energy. For the half-band width $D$ the value of one\\
is used.
\end{tabular}
\end{align*}

The fact that the three curves in Fig.3a are essentially parallel to each
other demonstrates that the density of states $\rho_{0}$ at the Fermi energy
is the appropriate parameter because otherwise the different curves would form
a finite angle. On the other hand the fact that the curves for the free
electrons and the tight binding are shifted with respect to the rectangular
band can be described by an effective half-band width $D^{\ast}$ which is
different from one. The shift is just equal to $\ln\left(  D^{\ast}\right)  $.
After adjusting an effective half-band width $D^{\ast}$ the plots of
$\ln\left(  \Delta E_{st}/D^{\ast}\right)  $ yield a universal curve for the
three cases as shown in Fig.3b.
\begin{align*}
&
{\includegraphics[
height=3.1681in,
width=3.6463in
]%
{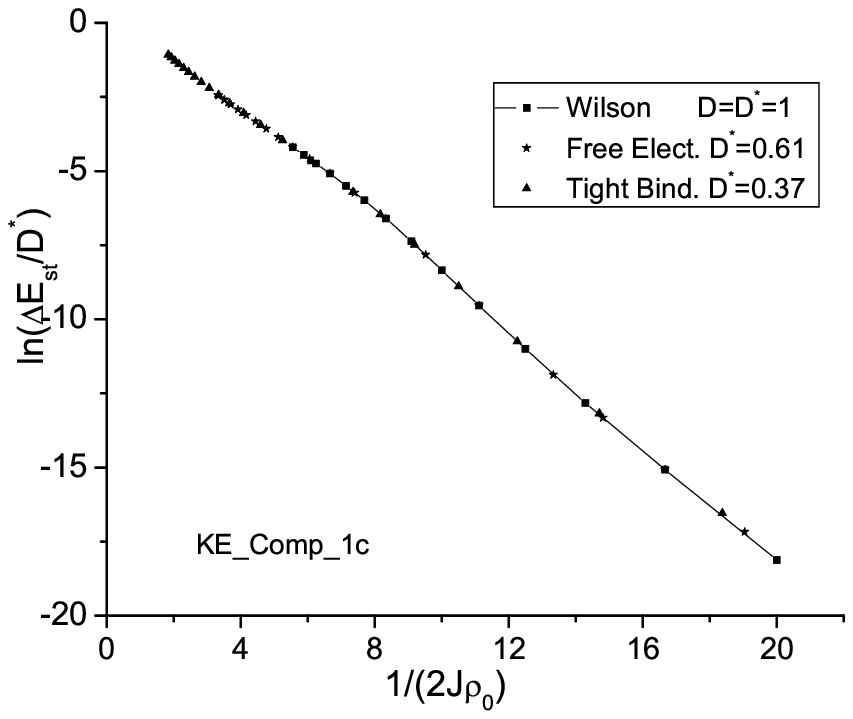}%
}%
\\
&
\begin{tabular}
[c]{l}%
Fig.3b: The same plot as in Fig.3a but using an effective\\
half-band width $D^{\ast}.$%
\end{tabular}
\end{align*}

In table I the effective values for the half-band width are collected in
column 3. In column 4 the effective band width $2D^{\ast}$ is multiplied with
$\rho_{0}$. The result is rather surprising. It yields the value one in all
three cases.

The Kondo energy is essentially given by
\[
k_{B}T_{K}\thickapprox D^{\ast}\exp\left(  -\frac{1}{2J\rho_{0}}\right)
\]
The density of states $\rho_{0}$ appears in denominator of the exponent while
the half-band width $D^{\ast}$ is just a pre-factor to the exponential
function. From the numerical calculation follows that for the very different
bands the product of $\rho_{0}$ and $2D^{\ast}$ is constant and equal to the
number of states (per spin and atom) in the whole band which is one. \
\begin{align*}
&
\begin{tabular}
[c]{|l|l|l|l|}\hline
\textbf{Spectrum} & $\mathbf{\rho}_{0}$ & $\mathbf{D}^{\ast}$ & $2\mathbf{D}%
^{\ast}\mathbf{\rho}_{0}$\\\hline
Wilson spectrum & 0.5 & 1 & 1\\\hline
Free Electrons & 0.75 & 0.66 & 0.98\\\hline
Tight Binding & 1.36 & .37 & 1.01\thinspace\\\hline
\end{tabular}
\ \ \\
&
\begin{tabular}
[c]{l}%
Table I: The density of states $\rho_{0}$ at the Fermi level,\\
the effective half-band width $D^{\ast}$ obtained from the\\
Kondo calculation and the product $2D^{\ast}\rho_{0}$ for bands\\
with very different band structure.
\end{tabular}
\end{align*}

\section{Conclusion}

The Wilson states are generalized to electron bands with energy dependent
density of states and s-d-interaction. These state represent a finite energy
range and possess the full interaction with the impurity. The Wilson states
permit us to introduce band-structure data into numerical calculations of the
Friedel-Anderson impurity and the Kondo impurity. This is demonstrated in
three examples. These calculations show that the properties of the singlet-
and triplet state of the Kondo impurity for an arbitrary band can be deduced
from the results for a rectangular Wilson band and the actual density of
states at the Fermi level.

Acknowledgement: The research was supported by the National Science Foundation
DMR-0439810.\newpage

\appendix

\section{Appendix}

\subsection{Friedel-Anderson impurity}

With the Wilson bases the Friedel-Anderson Hamiltonian has the form%
\begin{equation}
H_{FA}=\sum_{\sigma}\{\sum_{\nu=1}^{N}\varepsilon_{\nu}c_{\nu\sigma}^{\ast
}c_{\nu\sigma}+E_{d}d_{\sigma}^{\ast}d_{\sigma}+\sum_{\nu=1}^{N}V_{sd}%
(\nu)[d_{\sigma}^{\ast}c_{\nu\sigma}+c_{\nu\sigma}^{\ast}d_{\sigma}%
]\}+Un_{d+}n_{d-}%
\end{equation}

The author introduced a magnetic solution $\Psi_{MS}$ of the form%
\[
\Psi_{MS}=\left[  Aa_{0-\downarrow}^{\ast}a_{0+\uparrow}^{\ast}+Bd_{\downarrow
}^{\ast}a_{0+\uparrow}^{\ast}+Ca_{0-\downarrow}^{\ast}d_{\uparrow}^{\ast
}+Dd_{\downarrow}^{\ast}d_{\uparrow}^{\ast}\right]  \left\vert \mathbf{0}%
_{a+\uparrow}\mathbf{0}_{a-\downarrow}\right\rangle
\]
The states $a_{0+}^{\dag}$ and $a_{0-}^{\dag}$ are optimized localized states
formed from the Wilson states. They determine the full bases $\left\{
a_{i+}^{\dag}\right\}  $ and $\left\{  a_{i-}^{\dag}\right\}  $. The
many-electron state $\left\vert \mathbf{0}_{a+\uparrow}\mathbf{0}%
_{a-\downarrow}\right\rangle $ is given by $\prod_{i=1}^{n-1}a_{i+\uparrow
}^{\ast}\prod_{i=1}^{n-1}a_{i-\downarrow}^{\ast}\Phi_{0}$ where $\Phi_{0}$ is
the vacuum state.

This solution has a considerably lower energy than the mean-field solution and
yields lower magnetic moments than the mean-field solution. The threshold of
the Coulomb interaction $U$ \ to form a magnetic moment is almost twice as
large as in mean-field theory \cite{B152}.

However, the real ground state is a symmetric state. It is obtained by
reversing all spins in $\Psi_{MS}$ and superimposing the two solutions. Then
one obtains
\begin{equation}
\Psi_{SS}=\Psi_{MS}\left(  \uparrow\downarrow\right)  \mp\Psi_{MS}\left(
\downarrow\uparrow\right)  \label{PsiSS}%
\end{equation}%
\begin{align*}
&  =\left[  Aa_{0-\downarrow}^{\ast}a_{0+\uparrow}^{\ast}+Bd_{-\downarrow
}^{\ast}a_{0+\uparrow}^{\ast}+Ca_{0-\downarrow}^{\ast}d_{+\uparrow}^{\ast
}+Dd_{-\downarrow}^{\ast}d_{+\uparrow}^{\ast}\right]  \left\vert
\mathbf{0}_{a+\uparrow}\mathbf{0}_{a-\downarrow}\right\rangle \\
&  \mp\left[  \overline{A}a_{0-\uparrow}^{\ast}a_{0+\downarrow}^{\ast
}+\overline{B}d_{-\uparrow}^{\ast}a_{0+\downarrow}^{\ast}+\overline
{C}a_{0-\uparrow}^{\ast}d_{+\downarrow}^{\ast}+\overline{D}d_{-\uparrow}%
^{\ast}d_{+\downarrow}^{\ast}\right]  \left\vert \mathbf{0}_{a-\uparrow
}\mathbf{0}_{a+\downarrow}\right\rangle
\end{align*}

The final solution is obtained by minimizing the ground-state energy by
optimizing the two localized states $a_{0+}^{\dag}$ and $a_{0-}^{\dag}$. The
construction of the localized states $a_{0\pm}^{\dag}$ and the full basis
$\left\{  a_{i\pm}^{\dag}\right\}  $ as well as the optimization are described
in appendix A3.

This \ ground state is determined by the components of the two localized
states $a_{0+}^{\dag}$ and $a_{0-}^{\dag}$. Since we generally use 40 Wilson
states this requires 80 components. The resulting ground-state energy and the
occupation of the d-states is of the same accuracy as a large $\left(
1/N\right)  $ expansion up to second order which requires the calculation of
more than $10^{7}$ self-consistent amplitudes.

\subsection{Kondo wave-function}

The Kondo impurity has a magnetic moment. Therefore either the spin-up or
spin-down d-state is occupied. Zero or double occupancy of the d-states is
forbidden. Then the Coulomb interaction transforms into an exchange
interaction of the form%
\begin{equation}
H_{ex}=v_{a}J\left[
\begin{array}
[c]{c}%
\left(  S_{+}\Psi_{\downarrow}^{\dag}\left(  0\right)  \Psi_{\uparrow}\left(
0\right)  +S_{-}\Psi_{\uparrow}^{\dag}\left(  0\right)  \Psi_{\downarrow
}\left(  0\right)  \right) \\
+S_{z}\left(  \Psi_{\uparrow}^{\dag}\left(  0\right)  \Psi_{\uparrow}\left(
0\right)  -\Psi_{\downarrow}^{\dag}\left(  0\right)  \Psi_{\downarrow}\left(
0\right)  \right)
\end{array}
\right]  \label{Hsd}%
\end{equation}
where $S_{+},S_{-},S_{z}$ are the spin operators of the impurity with spin
$S=1/2$ and $\Psi_{\uparrow}^{\dag}\left(  0\right)  $ and $\Psi_{\downarrow
}^{\dag}\left(  0\right)  $ represent field operators.

In ref. \cite{B153} the approximate solution for the Kondo impurity was
introduced. It is given by equ. (\ref{Psi_K}) .
\begin{equation}
\psi_{K}=\left(  Ba_{0+\uparrow}^{\ast}d_{\downarrow}^{\ast}+Cd_{\uparrow
}^{\ast}a_{0-\downarrow}^{\ast}\right)  \left\vert \mathbf{0}_{a+\uparrow
}\mathbf{0}_{a-\downarrow}\right\rangle +\left(  \overline{C}a_{0-\uparrow
}^{\ast}d_{\downarrow}^{\ast}+\overline{B}d_{\uparrow}^{\ast}a_{0+\downarrow
}^{\ast}\right)  \left\vert \mathbf{0}_{a-\uparrow}\mathbf{0}_{a+\downarrow
}\right\rangle \label{Psi_K}%
\end{equation}
It uses two sets a bases, $\left\{  a_{+i}^{\dag}\right\}  $ and $\left\{
a_{i-}^{\dag}\right\}  $ which are orthogonal transformed versions of a Wilson
basis $\left\{  c_{\nu}^{\dag}\right\}  $. Again the many-electron state
$\left\vert \mathbf{0}_{a+\uparrow}\mathbf{0}_{a-\downarrow}\right\rangle $ is
given by $\prod_{i=1}^{n-1}a_{i+\uparrow}^{\ast}\prod_{i=1}^{n-1}%
a_{i-\downarrow}^{\ast}\Phi_{0}$. The first part of the ansatz consists of the
product of two $n$-electron state $\left\vert \mathbf{0}_{a+\uparrow
}\mathbf{0}_{a-\downarrow}\right\rangle $ multiplied with the two-particle
state $\left(  Ba_{0+\uparrow}^{\ast}d_{\downarrow}^{\ast}+Cd_{\uparrow}%
^{\ast}a_{0-\downarrow}^{\ast}\right)  .$ If one reverses all spins in this
term one obtains the second term (after rearranging the operators). This
solution is derived from the solution of the Friedel-Anderson impurity in the
limit where zero and double occupancy of the d-level is zero (for infinite
Coulomb potential $U$ and $E_{d}=-U/2$).

The final solution is obtained by minimizing the ground-state energy by
optimizing the two localized states $a_{0+}^{\dag}$ and $a_{0-}^{\dag}$. The
construction of the localized states $a_{0\pm}^{\dag}$ and the full basis
$\left\{  a_{i\pm}^{\dag}\right\}  $ as well as the optimization are described
in appendix A3.

If one sets $\overline{C}=C$ and $\overline{B}=B$ one obtains the singlet
ground state. If one permits a free variation of all four coefficients one
obtains besides the singlet ground state also the excited triplet state with
$\overline{C}=-C$ and $\overline{B}=-B$. We denote this state as the unrelaxed
triplet state. One obtains the relaxed triplet state when the opposite signs
of the coefficients are locked ($\overline{C}=-C$ and $\overline{B}=-B$). Then
the minimization of the\ energy yields the triplet state with the lowest
energy. In present paper the unrelaxed singlet-triplet excitation is plotted
for the different bands.

\subsection{Construction of the Basis $a_{0}^{\dag}$, $a_{i}^{\dag}$}

For the construction of the state $a_{0}^{\dag}$ and the rest of basis
$a_{i}^{\dag}$ one starts with the s-band electrons $\left\{  c_{\nu}^{\dag
}\right\}  $ which consist of $N$ states (for example Wilson's states). The
$d^{\dag}$-state is ignored for the moment. \newline

\begin{itemize}
\item In step (1) one forms a normalized state $a_{0}^{\dag}$ out of the
s-states with:
\end{itemize}

\begin{equation}
a_{0}^{\dag}=\sum_{\nu=1}^{N}\alpha_{\nu}^{0}c_{\nu}^{\dag}%
\end{equation}
The coefficients $\alpha_{\nu}^{0}$ can be arbitrary at first. One reasonable
choice is $\alpha_{\nu}^{0}=1/\sqrt{N}$

\begin{itemize}
\item In step (2) $\left(  N-1\right)  $ new basis states $a_{i}^{\dag}$
$\left(  1\leq i\leq N-1\right)  $ are formed which are normalized and
orthogonal to each other and to $a_{0}^{\dag}$.

\item In step (3) the s-band Hamiltonian $H_{0}$ is constructed in this new
basis. One puts the state $a_{0}^{\dag}$ at the top so that its matrix
elements are $H_{0i}$ and $H_{i0}$.

\item In step (4) the $\left(  N-1\right)  $-sub Hamiltonian which does not
contain the state $a_{0}^{\dag}$ is diagonalized. The resulting Hamilton
matrix for the s-band then has the form%
\begin{equation}
H_{0}=\left(
\begin{array}
[c]{ccccc}%
E(0) & V_{fr}(1) & V_{fr}(2) & ... & V_{fr}(N-1)\\
V_{fr}(1) & E(1) & 0 & ... & 0\\
V_{fr}(2) & 0 & E(2) & ... & 0\\
.. & ... & ... & ... & ...\\
V_{fr}(N-1) & 0 & 0 & ... & E(N-1)
\end{array}
\right)  \label{hmat}%
\end{equation}
The creation operators of the new basis are given by a new set of $\left\{
a_{i}^{\dag}\right\}  ,$ ($0<i\leq N-1)$. Again the $a_{i}^{\dag}$ can be
expressed in terms of the s-states; $a_{i}^{\dag}=\sum_{\nu=1}^{N}\alpha_{\nu
}^{i}c_{\nu}^{\dag}$. After the state $a_{0}^{\dag}$ is constructed the other
states $a_{i}^{\dag}$ are uniquely determined. The additional s-d hopping
Hamiltonian can be expressed in the terms of the new basis, and one obtains
the Friedel Hamiltonian as given in eq. (\ref{hfr'}). The state $\Psi_{SS}$ is
formed, and the energy expectation value (of the full Hamiltonian) is calculated.

\item In the final step (5) the state $a_{0}^{\dag\text{ }}$is rotated in the
$N$-dimensional Hilbert space. In each cycle the state $a_{0}^{\dag}$ is
rotated in the $\left(  a_{0}^{\dag}\text{,}a_{i_{0}}^{\dag}\right)  $ plane
by an angle $\theta_{i_{0}}$ for $1\leq i_{0}\leq N-1$. Each rotation by
$\theta_{i_{0}}$ yields a new $\overline{a_{0}}^{\dag}$
\[
\overline{a_{0}}^{\dag}=a_{0}^{\dag}\cos\theta_{i_{0}}+a_{i_{0}}^{\dag}%
\sin\theta_{i_{0}}%
\]

The rotation leaves the whole basis $\left\{  a_{0}^{\dag},a_{i}^{\dag
}\right\}  $ orthonormal. Step (4), the diagonalization of the $\left(
N-1\right)  $-sub Hamiltonian, is now much quicker because the $\left(
N-1\right)  $-sub-Hamiltonian is already diagonal with the exception of the
$i_{0}$- row and the $i_{0}$-column . For each rotation plane $\left(
a_{0}^{\dag}\text{,}a_{i_{0}}^{\dag}\right)  $ the optimal $a_{0}^{\dag}$ with
the lowest energy expectation value is determined. This cycle is repeated
until one reaches the absolute minimum of the energy expectation value. In the
example of the Friedel resonance Hamiltonian this energy agrees numerically
with an accuracy of $10^{-15}$ with the exact ground-state energy of the
Friedel Hamiltonian \cite{B91}. For the Friedel-Anderson impurity the
procedure is stopped when the expectation value changes by less than
$10^{-10}$ during a full cycle.
\end{itemize}

\end{document}